\documentclass{akws-procs9x6}

\def\al{\alpha}
\def\be{\beta}
\def\ga{\gamma}
\def\de{\delta}
\def\ep{\epsilon}

\def\ka{\kappa}
\def\la{\lambda}

\def\si{\sigma}

\def\ps{\psi}
\def\om{\omega}
\def\Ga{\Gamma}

\def\La{\Lambda}

\def\cl{{\mathcal L}}

\def\fr#1#2{{{#1} \over {#2}}}
\def\half{{\textstyle{1\over 2}}}

\def\frac#1#2{{\textstyle{{#1}\over {#2}}}}

\def\lsim{\mathrel{\rlap{\lower4pt\hbox{\hskip1pt$\sim$}}
    \raise1pt\hbox{$<$}}}
\def\gsim{\mathrel{\rlap{\lower4pt\hbox{\hskip1pt$\sim$}}
    \raise1pt\hbox{$>$}}}
\def\sqr#1#2{{\vcenter{\vbox{\hrule height.#2pt
         \hbox{\vrule width.#2pt height#1pt \kern#1pt
         \vrule width.#2pt}
         \hrule height.#2pt}}}}

\def\prt{\partial}
\def\lrpartial{\raise 1pt\hbox{$\stackrel\leftrightarrow\partial$}}

\def\lrDmu{\stackrel{\leftrightarrow}{D_\mu}}

\def\etal{{\it et al.}}

\def\pt#1{\phantom{#1}}

\def\ol#1{\overline{#1}}

\def\a{$a_\mu$}
\def\b{$b_\mu$}
\def\c{$c_{\mu\nu}$}
\def\d{$d_{\mu\nu}$}
\def\e{$e_\mu$}
\def\f{$f_\mu$}
\def\g{$g_{\la\mu\nu}$}
\def\H{$H_{\mu\nu}$}
\def\kaa{$(k_{A})_\mu$}
\def\kaf{$(k_{AF})_\mu$}
\def\kf{$(k_{F})_{\ka\la\mu\nu}$}

\def\tt{$t^{\ka\la\mu\nu}$}

\def\lrDmu{{\hskip -3 pt}\stackrel{\leftrightarrow}{D_\mu}{\hskip -2pt}}

\def\nsc#1#2#3{\om_{#1}^{{\pt{#1}}#2#3}}

\def\tor#1#2#3{T^{#1}_{{\pt{#1}}#2#3}}

\def\vb#1#2{e_{#1}^{{\pt{#1}}#2}}
\def\ivb#1#2{e^{#1}_{{\pt{#1}}#2}}
\def\uvb#1#2{e^{#1#2}}
\def\lvb#1#2{e_{#1#2}}

\newcommand{\beq}{\begin{equation}}
\newcommand{\eeq}{\end{equation}}
\newcommand{\bea}{\begin{eqnarray}}
\newcommand{\eea}{\end{eqnarray}}
\newcommand{\bit}{\begin{itemize}}
\newcommand{\eit}{\end{itemize}}
\newcommand{\rf}[1]{(\ref{#1})}

\begin{document}

\title{Lorentz Violation and Gravity}

\author{V.\ ALAN KOSTELECK\'Y}

\address{
Physics Department, Indiana University \\
Bloomington, IN 47405, U.S.A.}

\maketitle

\abstracts{
Lorentz symmetry lies at the heart of relativity
and is a feature of low-energy descriptions of nature.
Minuscule Lorentz-violating effects
arising in theories of quantum gravity
offer a promising candidate signal for new physics
at the Planck scale.
A framework is presented for
incorporating Lorentz violation
into general relativity and other theories of gravity.
Applying this framework yields
a proof that explicit Lorentz symmetry breaking
is incompatible with generic Riemann-Cartan geometries.
The framework also enables the construction of all possible terms 
in the effective low-energy action for the underlying quantum gravity.
These terms form 
the gravitationally coupled Standard-Model Extension (SME),
which offers a comprehensive guide to searches
for observable phenomena.
The dominant and sub-dominant Lorentz-violating terms 
in the gravitational and QED sectors of the SME are discussed.
}

\section{Introduction}

Reconciling gravity with quantum mechanics
to form a consistent theory of quantum gravity
remains a major outstanding problem in theoretical physics. 
The difficulty of the problem 
is exacerbated by the lack of experimental guidance. 
Progress in physics is often made
through the combination of theory and experiment working in tandem,
but the natural scale for quantum gravity is the Planck scale,
which lies some 17 orders of magnitude above our 
presently attainable energy scales.
At first sight,
this appears an insuperable barrier
to the acquisition of experimental information
about quantum gravity.

Remarkably,
under suitable circumstances,
some experimental information about quantum gravity
can nonetheless be obtained.
The point is that minuscule effects emerging 
from the underlying quantum gravity 
might be detected in sufficiently sensitive experiments.
To be identified as definitive signals from the Planck scale,
such effects would need to violate 
some established principle of low-energy physics.
One promising class of potential effects is relativity violations,
arising from breaking the Lorentz symmetry 
that lies at the heart of relativity.\cite{kps}
Recent proposals suggest these effects could emerge from 
strings, loop quantum gravity, 
noncommutative field theories, or numerous other sources
at the Planck scale.\cite{confs} 

Whatever the nature of the underlying quantum gravity,
effective field theory is an appropriate tool
for the general description of low-energy signals 
of Lorentz violation.\cite{kpo}
To be realistic,
a theory of this type must reproduce established physics.
In Minkowski spacetime,
nongravitational phenomena involving 
the basic particles and forces down to the quantum level
are successfully described by the Standard Model (SM)
of particle physics.
Adding gravitational couplings and the Einstein-Hilbert action
for general relativity yields 
the gravitationally coupled SM,
which encompasses all known fundamental physics.
This combined theory must therefore be a basic component 
of any realistic effective field theory.

In Minkowski spacetime,
relativity violations can be incorporated 
as additional terms in the SM action 
describing arbitrary coordinate-independent Lorentz violation,
and all dominant contributions 
at low energies are explicitly known.\cite{ck}
However,
the inclusion of Lorentz violation in an effective field theory 
containing also the Einstein-Hilbert action 
and the gravitationally coupled SM is more challenging.
The study of relativity violations 
in the corresponding spacetimes requires a framework 
allowing violations of local Lorentz invariance
while preserving general coordinate invariance.
Also,
since physical matter is formed from leptons and quarks,
the framework must be sufficiently supple to incorporate spinors.

This talk summarizes a suitable framework 
that meets all the above criteria,
along with some key associated results.
The framework described enables the construction of the general 
low-energy effective field theory,
the Standard-Model Extension (SME),
which serves as a comprehensive basis for 
theoretical and experimental studies of Lorentz violation
in all gravitational and SM sectors.
The talk is based on a selection of results 
obtained in Ref.\ \refcite{akgrav},
to which the reader is referred for more details.

\section{Framework}

The framework summarized here,
appropriate for the comprehensive description of Lorentz violation,
is founded on Riemann-Cartan geometry 
and the vierbein formalism.\cite{uk}
This formalism naturally distinguishes 
local Lorentz and general coordinate transformations
and also allows the treatment of spinors.
The basic gravitational fields 
are the vierbein $\vb \mu a$
and the spin connection $\nsc \mu a b$,
and the action of the local Lorentz group
at each spacetime point
allows three rotations and three boosts
independent of general coordinate transformations.
In this context,
Lorentz violation appears in a local Lorentz frame 
when a nonzero vacuum value exists for one or more quantities 
carrying local Lorentz indices,
called coefficients for Lorentz violation.

As an illustrative example for the basic ideas 
of the framework,
suppose a nonzero timelike coefficient $k_a = (k,0,0,0)$
exists in a certain local Lorentz frame at some point $P$.
Whenever particles (or localized fields) 
have observable interactions with $k_a$,
physical Lorentz violation occurs.
The corresponding Lorentz transformations,
called local {\it particle} Lorentz transformations,
act to boost or rotate particles in the fixed local frame at $P$, 
leaving $k_a$ and any other background quantities unaffected.
Note, however, that the local Lorentz frame itself can be 
changed by local {\it observer} Lorentz transformations,
under which $k_a$ behaves covariantly as a four-vector.
Note also that the physics is covariant 
under general coordinate transformations,
as desired,
because a change of the observer's spacetime coordinates $x^\mu$
induces a conventional general coordinate transformation 
on $k_\mu = \vb \mu a k_a$.
The breaking of Lorentz symmetry is called explicit
if $k_\mu (x) $ is specified as a predetermined external quantity,
while it is spontaneous 
if instead $k_\mu (x)$ is determined
through a dynamical procedure such as 
the development of a vacuum value.

In general,
the Lorentz-violating piece $S_{\rm LV}$ 
of the action for the effective field theory 
contains terms of the form
\beq
S_{\rm LV} \supset 
\int d^4 x ~ e k_x J^x ,
\label{slv}
\eeq
where 
$k_x$ is a coefficient for Lorentz violation
in the covariant $x$ representation of the observer Lorentz group.
Also,
$J^x$ lies in the corresponding contravariant representation
and is a general-coordinate invariant
formed from the vierbein, spin connection, and SM fields.
The form \rf{slv}
of terms in the effective action
is independent of the origin of the Lorentz violation
in the underlying quantum gravity,
including whether the violation is spontaneous or explicit.

\section{Spontaneous and Explicit Lorentz Violation}

With this framework established,
various issues concerning observable Lorentz violation 
can be addressed.
One result is that
explicit and spontaneous Lorentz violations
have distinct implications
for the energy-momentum tensor.
To see this,
first separate the action of the full effective theory 
into a piece $S_{\rm gravity}$
involving only the vierbein and spin connection 
and the remainder,
$S_{\rm matter} = S_{\rm matter, 0} + S_{\rm matter, LV}$,
where $S_{\rm matter, LV}$ contains 
all Lorentz violations involving matter.
Any term in the latter therefore has the general form \rf{slv},
where the operator $J^x$ is now taken 
to be formed from matter fields $f^y$ 
and their covariant derivatives.

For {\it explicit} Lorentz violation,
consider a particular variation of $S_{\rm matter}$
for which all fields and backgrounds are allowed to vary,
including the coefficients for explicit Lorentz violation,
but in which the dynamical fields $f^y$
satisfy equations of motion:
\bea
\de S_{\rm matter}
&=& 
\int d^4 x ~ e ({T}^{\mu\nu} \lvb \nu a \de \vb \mu a
+ \half {S}^\mu_{\pt{\mu}ab} \de \nsc \mu a b
+ e J^x\de k_x).
\label{svar}
\eea
This expression defines the energy-momentum tensor ${T}^{\mu\nu}$
and the spin-density tensor ${S}^\mu_{\pt{\mu}ab}$,
as usual.
For infinitesimal local Lorentz transformations
$\de \vb \mu a$, $\de \nsc \mu a b$, $\de k_x$, 
the variation \rf{svar} yields the condition
\bea
&&{T}^{\mu\nu} - {T}^{\nu\mu} - 
(D_\al - T^\be_{\pt{\be}\be\al}) {S}^{\al\mu\nu}
= - \uvb \mu a \uvb \nu b k_x (X_{[ab]})^x_{\pt{x}y} J^y
\label{emsym}
\eea
on the symmetry of the energy-momentum tensor ${T}^{\mu\nu}$,
where $D_\mu$ is the covariant derivative,
$T^\la_{\pt{\la}\mu\nu}$ is the torsion,
and 
$(X_{[ab]})^x_{\pt{x}y}$ is the representation for the 
local Lorentz generators. 
When instead the special variation \rf{svar}
is induced by a diffeomorphism,
the variations $\de \vb \mu a$, $\de \nsc \mu a b$, $\de k_x$
are Lie derivatives,
yielding the covariant conservation law 
\bea
&&
(D_\mu - T^\la_{\pt{\la}\la\mu}) {T}^\mu_{\pt{\mu}\nu}
+ T^\la_{\pt{\la}\mu\nu} {T}^\mu_{\pt{\mu}\la}
+ \half R^{ab}_{\pt{ab}\mu\nu} {S}^\mu_{\pt{\mu}ab}
= J^x D_\nu k_x  ,
\label{emcons}
\eea
where $R^{ab}_{\pt{ab}\mu\nu}$ is the curvature.
In the limit of conventional general relativity,
these equations reduce to the familiar expressions 
${T}^{\mu\nu}={T}^{\nu\mu}$ and 
$D_\mu {T}^\mu_{\pt{\mu}\nu} = 0$.
In the Minkowski-spacetime limit with Lorentz violation,
known results\cite{ck}
also emerge.

For {\it spontaneous} Lorentz violation,
the derivation can be adapted to obtain 
equations similar to \rf{emsym} and \rf{emcons},
but with the terms involving $k_x$ replaced by zero. 
This is because all coefficients arising from spontaneous breaking
are vacuum field values
and therefore must obey equations of motion,
so the variation $\de k_x$ in Eq.\ \rf{svar} is absent.
The result can also be understood geometrically.
The spacetime geometry 
implies a set of identities, the Bianchi identities,
that are tied to the equations of motion
and hence imply certain conditions on the matter sources.
However,
for sources involving explicit Lorentz violation,
these conditions are generically incompatible 
with covariant conservation laws for the matter.
For example,
in general relativity
the Bianchi identities are $D_\mu G^{\mu\nu} = 0$,
the Einstein equations are $G^{\mu\nu} = 8 \pi G_N T^{\mu\nu}$,
and substitution yields the condition $D_\mu T^{\mu\nu} = 0$,
which in the presence of explicit Lorentz violation
is incompatible with the result \rf{emcons}.
In contrast,
spontaneous Lorentz violation yields consistent results,
essentially because 
in this case the coefficients for Lorentz violation
form an intrinsic part of the geometrical structure
rather than being externally imposed.

\section{Low-Energy Effective Action}

A wide-ranging application of the general framework 
summarized here 
is the construction of all possible dominant terms 
in the low-energy effective action,
independent of the structure 
of the underlying quantum gravity theory.
The full SME effective action at low energies 
is a sum of partial actions,
\beq
S_{\rm SME} = S_{\rm gravity} + S_{\rm SM} + S_{\rm LV} + \ldots .
\eeq
Here,
the term $S_{\rm gravity}$ represents the pure-gravity sector,
involving the vierbein and the spin connection
and including any Lorentz violation.
The term $S_{\rm SM}$ is the SM action
with gravitational couplings.
The term $S_{\rm LV}$
contains all Lorentz-violating terms 
that involve matter fields and dominate at low energies,
including minimal gravitational couplings.
The ellipsis represents low-energy terms of higher suppression order,
including operators of mass dimension greater than four,
some of which violate Lorentz symmetry.

The pure-gravity action can be written 
\beq
S_{\rm gravity} = 
\fr 1 {16\pi G_N }
\int d^4 x ~ 
({\cl}_{e, \om}^{\rm LI}
+{\cl}_{e, \om}^{\rm LV}
+\ldots ),
\label{gravact}
\eeq
where the Lorentz-invariant piece ${\cl}_{e, \om}^{\rm LI}$ 
and the Lorentz-violating piece ${\cl}_{e, \om}^{\rm LV}$
involve only $\vb \mu a$ and $\nsc \mu a b$.
The ellipsis represents possible dependence
on nonminimal dynamical gravitational fields,
such as the recently proposed cosmologically varying scalar fields
that can lead to Lorentz violation.\cite{klpe}
The Lorentz-invariant lagrangian 
${\cl}_{e, \om}^{\rm LI}$ 
can be expanded as usual,
\beq
{\cl}_{e, \om}^{\rm LI} = eR -2e\La + \ldots ,
\label{lilag}
\eeq
while the Lorentz-violating lagrangian 
${\cl}_{e, \om}^{\rm LV}$
has the form
\bea
{\cl}_{e, \om}^{\rm LV}
&=&
e (k_T)^{\la\mu\nu} T_{\la\mu\nu}
+ e (k_R)^{\ka\la\mu\nu} R_{\ka\la\mu\nu}
\nonumber\\
&&
+ e (k_{TT})^{\al\be\ga\la\mu\nu} T_{\al\be\ga} T_{\la\mu\nu}
+\ldots .
\label{lvlag}
\eea

The Lorentz-violating matter action $S_{\rm LV}$
can also be constructed as a series
of terms involving both SM and gravitational fields.
For illustrative purposes,
attention here is restricted to the special limit of 
single-fermion gravitationally coupled quantum electrodynamics (QED),
for which only the dominant and minimally coupled terms 
are considered. 
A discussion of the full theory can be found 
in Ref.\ \refcite{akgrav}.

In this limit, 
the relevant U(1)-invariant action is a sum of partial actions 
for the Dirac fermion $\ps$ and the photon $A_\mu$.
The fermion partial action for the QED extension can be written as
\beq
S_\ps
=
\int d^4 x
(\half i e \ivb \mu a \ol \ps \Ga^a \lrDmu \ps
- e \ol \ps M \ps) ,
\label{qedxps}
\eeq
where the symbols $\Ga^a$ and $M$
are defined by
\bea
\Ga^a
&\equiv &
\ga^a
- c_{\mu\nu} \uvb \nu a \ivb \mu b \ga^b
- d_{\mu\nu} \uvb \nu a \ivb \mu b \ga_5 \ga^b
\nonumber\\
&&
- e_\mu \uvb \mu a
- i f_\mu \uvb \mu a \ga_5
- \half g_{\la\mu\nu} \uvb \nu a \ivb \la b \ivb \mu c \si^{bc},
\label{gamdef}
\eea
\bea
M
&\equiv &
m
+ i m_5 \ga_5
+ a_\mu \ivb \mu a \ga^a
+ b_\mu \ivb \mu a \ga_5 \ga^a
+ \half H_{\mu\nu} \ivb \mu a \ivb \nu b \si^{ab} .
\label{mdef}
\eea
The first term of Eq.\ \rf{gamdef} 
and the first two terms of Eq.\ \rf{mdef}
are conventional,
while the others involve Lorentz violation 
controlled by the coefficients 
\a, \b, \c, \d, \e, \f, \g, \H,
which typically vary with position.
The covariant derivative $D_\mu$ in Eq.\ \rf{qedxps} 
is a combination of the spacetime covariant derivative
and the usual U(1) covariant derivative:
\beq
D_\mu \ps \equiv
\prt_\mu \ps
+ \frac 14 i \nsc \mu a b \si_{ab} \ps
- i q A_\mu \ps .
\label{covderivqed}
\eeq

In the photon sector, 
the partial action is 
\beq
S_A
=
\int d^4 x
({\cl}_{F} + {\cl}_{A}),
\label{qedxph}
\eeq
where
\bea
{\cl}_{F}
&=&
-\frac 14 e F_{\mu\nu}F^{\mu\nu}
-\frac 14 e (k_F)_{\ka\la\mu\nu} F^{\ka\la} F^{\mu\nu},
\label{photlageven}\\
{\cl}_{A}
&=&
\half e (k_{AF})^\ka \ep_{\ka\la\mu\nu} A^\la F^{\mu\nu}
- e (k_A)_\ka A^\ka.
\label{photlagodd}
\eea
The electromagnetic field strength $F_{\mu\nu}$
is defined by the locally U(1)-invariant form
\bea
F_{\mu\nu}
&\equiv&
D_\mu A_\nu - D_\nu A_\mu
+ \tor \la \mu \nu A_\la
\prt_\mu A_\nu - \prt_\nu A_\mu .
\label{fieldstr}
\eea
The Lorentz violation in this sector is controlled
by the coefficients \kf, \kaf, and \kaa. 

In Minkowski spacetime,
the coefficients for Lorentz violation in the SME predict a plethora 
of experimental signals for relativity violations,
even when attention is limited to
spacetime-constant coefficients 
for operators of mass dimension four or less.
Experimental tests in this limit to date include
ones with
photons,\cite{photonexpt,photonth}
electrons,\cite{eexpt,eexpt2,eexpt3}
protons and neutrons,\cite{ccexpt,spaceexpt}
mesons,\cite{hadronexpt}
muons,\cite{muexpt}
neutrinos,\cite{nuexpt,nuth}
and the Higgs.\cite{higgs}

In the full SME effective action including the gravitational couplings,
the Lorentz-violating terms 
create spacetime anisotropies and spacetime-dependent rescalings
of the coupling constants in the field equations,
which in turn induce further potentially significant physical effects.
The Lorentz-violating behaviors
of gravity modes and fundamental particles 
vary with momentum magnitude and orientation,
spin magnitude and orientation, 
and particle species. 
Established results for post-newtonian physics,
gravitational waves, black holes, cosmologies, 
and other standard scenarios 
typically acquire corrections 
depending on coefficients for Lorentz violation.

In the gravitational sector,
substantial deviations from conventional physics
due to Lorentz violation are likely  
only in regions of large gravitational fields,
such as near black holes or in the early Universe.
Nonetheless,
observable effects may emerge under suitable circumstances.
For example,
searches for Lorentz violation are feasible  
in laboratory and space-based experiments 
studying post-newtonian gravitational physics,\cite{cmw}
including the classic tests of gravitational physics,
of the inverse square law,
and of gravitomagnetic effects.
Similarly,
spacetime anisotropies in the equations for gravitational waves\cite{ks}
can be sought in Earth- or space-based experiments.
Comparisons of the speeds of neutrinos, light,
and gravitational waves
which can differ in the presence of Lorentz violation,
may also eventually be feasible by observing certain 
violent astrophysical processes.
On a larger scale,
anisotropic Lorentz-violating corrections generated for the 
conventional homogeneous FRW cosmologies
have the potential to generate a realistic anisotropic cosmology
with detectable effects.
One possible class of Lorentz-violating cosmological signals
would be alignment anomalies on large angular scales,
which have been reported in the WMAP data\cite{wmap}
but are absent in standard cosmologies.\cite{otzh}
Certain coefficients for Lorentz violation 
can also contribute to an effective cosmological constant,
dark matter, and dark energy.
For instance,
the small nonzero cosmological constant 
may be partially or entirely tied to small Lorentz violation 
and may also vary with spacetime position.

\section{Summary}

The gravitationally coupled SME discussed in this talk
is the full low-energy effective field theory
for gravitation and other fundamental interactions.\cite{akgrav} 
It offers a comprehensive basis for the study 
and analysis of experimental tests of Lorentz symmetry,
independent of the underlying quantum gravity. 
The detailed exploration of the associated
theoretical and experimental implications  
is an open challenge of considerable interest,
with the potential to uncover experimental signals 
from the underlying Planck-scale theory.

\section*{Acknowledgments}
This work was supported in part 
by NASA grants NAG8-1770 and NAG3-2194
and by DoE grant DE-FG02-91ER40661.

\end{document}